\documentclass[twocolumn,showpacs,preprintnumbers,amsmath,amssymb]{revtex4-1}

\usepackage{graphicx}
\usepackage{dcolumn}
\usepackage{bm}

\begin{document}

\title{Piezoelectric rotator for studying quantum effects in semiconductor nanostructures at high magnetic fields and low temperatures}

\author{L. A. Yeoh, A. Srinivasan}
\author{T. P. Martin}
\email{Now at: Acoustics Division, Naval Research Laboratory, Washington, DC 20375, United States of America}
\author{O. Klochan, A. P. Micolich}
\author{A. R. Hamilton}
\email{Alex.Hamilton@unsw.edu.au}
\affiliation{School of Physics, University of New South Wales, Sydney NSW 2052, Australia}

\date{\today}

\begin{abstract}
We report the design and development of a piezoelectric sample rotation system, and its integration into an Oxford Instruments Kelvinox $100$ dilution refrigerator, for orientation-dependent studies of quantum transport in semiconductor nanodevices at millikelvin temperatures in magnetic fields up to $10$~T. Our apparatus allows for continuous {\it in situ} rotation of a device through $>100^{\circ}$ in two possible configurations. The first enables rotation of the field within the plane of the device, and the second allows the field to be rotated from in-plane to perpendicular to the device plane. An integrated angle sensor coupled with a closed-loop feedback system allows the device orientation to be known to within $\pm 0.03^{\circ}$ whilst maintaining the sample temperature below $100$~mK.
\end{abstract}

\maketitle

\section{\label{sec:intro}INTRODUCTION}

There is growing interest in the development of spintronic (spin-electronic) devices that use the spin of an electron rather than its charge to represent and carry information~\cite{WolfSci01}. In particular, recent research has focused upon the manipulation and control of individual spins using electric rather than magnetic fields via the spin-orbit interaction~\cite{AwschalomPhys09}. The combined effects of the strong spin-orbit interaction and spatial confinement of charge carriers in a semiconductor heterostructure with broken inversion symmetry can lead to rather complex spin-physics~\cite{WinklerBook03}. A key manifestation of this complexity is in the strongly anisotropic Zeeman spin-splitting observed in hole quantum wires in AlGaAs/GaAs heterostructures~\cite{DanneauPRL06, KlochanNJP09, ChenNJP10}. In order to study this anisotropic Zeeman spin-splitting  more fully, it is highly desirable to have precise angular control and read-out during {\it in situ} rotation of the device with respect to an applied magnetic field, while at the same time  maintaining the device at the ultra-low temperatures required for electrical measurement of its quantum transport properties. This capability is also of more widespread interest in physics, facilitating low temperature magnetic field orientation dependence studies of, for example, commensurability effects in organic superconductors~\cite{BoebingerPRL90, OsadaPRL91, StorrPRB91}, critical field anisotropy in cuprate~\cite{RaffyPRL91, HoferPRB00} and ruthenate~\cite{MaoPRL00} superconductors, spin transitions and ferromagnetic ordering in quantum Hall systems~\cite{EisensteinPRB90, SchmellerPRL95, MurakiPRB99, KumadaPRL02}, bistability in hybrid semiconductor/ferromagnet systems~\cite{MeierJMMM00}, polarization reversal in multiferroics~\cite{AbePRL07} and studies of novel molecular magnet systems~\cite{MansonCM08}.

Designing such a rotation system is a challenge due to the numerous constraints involved in operating at millikelvin temperatures, under vacuum and within the small space available inside the bore of a high-field superconducting magnet. A number of rotation mechanisms developed for low temperature, high magnetic field operation have been previously reported in the literature. These were mostly based upon a mechanical gear mechanism, such as that described by Bhattacharya {\it et al.} who used a worm-gear mechanism driven by an external stepper motor to perform rotations at a temperature of $2$~K inside a Quantum Design MPMS SQUID susceptometer system~\cite{BhattacharyaRSI98}. However, such an approach is not practical for our applications for two reasons. First, in order  to pass into the inner vacuum chamber of the dilution refrigerator, this design would require a hermetically sealed drive shaft that can operate at a temperature $\sim4$~K. Second, a considerable amount of heat is generated due to friction, leading to a large jump in temperature when rotation is actuated. Palm and Murphy showed that this friction problem can be overcome to some extent by using sapphire vee jewels and matching pivots~\cite{PalmRSI99}. They achieved a power dissipation as low as $70~\mu$W for slow rotation rates of $0.33~^{\circ}$/s. This gave a maximum temperature rise of $8$~mK at the $30$~mK base temperature of their Oxford TLM-400 dilution refrigerator, which had a cooling power of $500~\mu$W at $T = 100$~mK. However, this design still requires a mechanical connection between room temperature and the sample space.

Piezoelectric actuators provide an alternative method for achieving rotation that eliminates the need for a mechanical drive shaft, as demonstrated for room temperature operation in an ultrahigh vacuum environment by de Haas {\it et al}.~\cite{deHaasRSI96}. One advantage of piezoelectric rotators is that they can be controlled electrically, resulting in easy automation of the rotator via an external controller unit. Although the wires used for control and readout also require hermetic feedthroughs for use inside the inner vacuum can of a dilution refrigerator, they provide a more elegant solution, as the design is free from mechanical sealing problems, additional heat leakage via thermal conduction through the drive shaft and frictional dissipation from gearing during rotation. Ohmichi {\it et al.}  demonstrated piezoactuated rotation at temperatures below $100$~mK and magnetic fields up to $33$~T in an Oxford TLM-100 with an estimated dissipation under $100~\mu$W at a rotation rate of $0.025~^{\circ}$/s with the rotator and sample immersed in the $^{3}$He-$^{4}$He mixture during operation~\cite{OhmichiRSI01}. However, the rotator design reported by Ohmichi {\it et al.} was limited to rotations of $\pm 10^{\circ}$ at most, albeit with a high angular resolution of $\sim~0.01^{\circ}$. For studying the Zeeman spin-splitting anisotropy in quantum devices, a much larger angular range of at least $90^{\circ}$ is required, which was a central focus of the rotation system reported here. We also required operation in vacuum, with the rotator mounted on a cold finger inside the bore of the superconducting magnet, placing even more stringent requirements on the power dissipation during device rotation.

\section{\label{sec:rotsys}DESIGN AND CONSTRUCTION}

There were a number of design requirements involved in producing a low temperature, high field rotation system that could meet the challenges involved in studying anisotropic spin effects in quantum devices. These included:

\begin{itemize}
\item The ability to achieve a sample rotation range $>~90^{\circ}$ between parallel and perpendicular orientations at minimum.
\item The ability to rotate a standard 20-pin leadless chip carrier (LCC20) with twenty sample wires in two separate configurations, with respect to the external magnetic field.
\item Achieve an angle readout resolution better than $0.1^{\circ}$.
\item Heat dissipation kept to a minimum ($<1$~mW) with minimal temperature fluctuation during measurement.
\item The rotator should consist of non-magnetic components, so that it can operate accurately at high magnetic fields up to $15$~T.
\item The ability to fit within the $39$~mm diameter sample space available inside the refrigerator vacuum cans and bore of the superconducting magnet.
\end{itemize}

\noindent{\bf The Rotator Assembly:} Our devices are fabricated using standard semiconductor device processing techniques on AlGaAs/GaAs heterostructures and are mounted in ceramic LCC $20$ device packages~\cite{SpectrumLCC20} for measurement. Two sample mounting configurations were designed, as depicted in Fig.~1, allowing for the sample to be tilted in two different orientations. In the first configuration, shown in Fig.~1(a), the rotation axis is aligned normal to the semiconductor heterostructure (i.e., parallel to the growth direction) allowing the magnetic field to be rotated in the plane of the heterostructure. In the second configuration, shown in Fig.~1(b), the rotation axis is aligned in the plane of the heterostructure allowing the magnetic field to be rotated from within the plane of the heterostructure, to normal to the heterostructure and directions in between.

\begin{figure}
\includegraphics[width=0.99\linewidth]{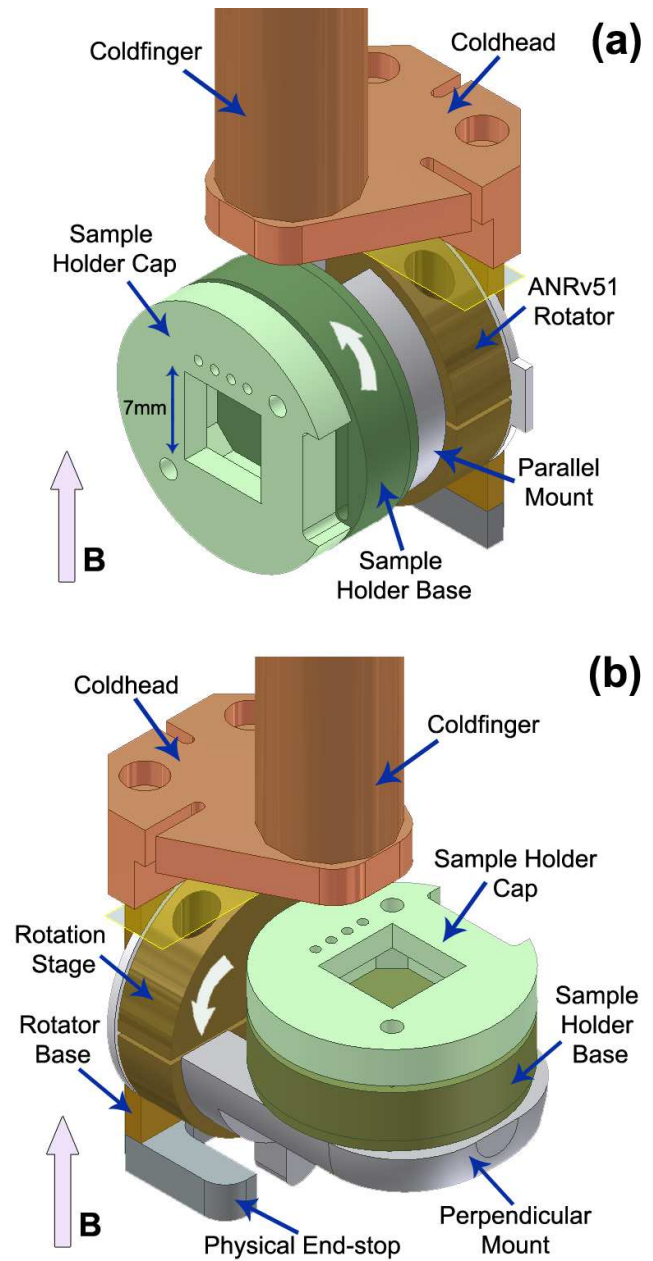}
\caption{Rotation system assembly for rotating the sample in two separate configurations with respect to the applied magnetic field $B$: (a) In-plane rotation with respect to field, (b) Out-of-plane field rotation.}
\end{figure}

Our design was based upon the commercially available piezoelectric ANRv$51$/RES nano-rotation stage from Attocube Systems AG~\cite{AttoRot09}, which features a built-in resistance-based electronic angle readout mechanism. The rotator is controlled by an external Attocube ANC$350$ controller unit~\cite{AttoRot09}, which allows for the automation of the experiment using customized LabVIEW procedures. The inertial drive principle is used to achieve rotation, which relies upon the combined effects of friction and inertia to achieve a net displacement. A saw-tooth driving voltage is used to power the rotator, with amplitudes ranging from $20-70$~V at frequencies up to $1$~kHz. Due to the reliance on a frictional `slip-stick' driving mechanism, the rotator has a limited torque, which is further reduced at low temperatures. To account for this, the sample mounts were designed with a small mass to minimize both the load on the rotator and accommodate the sample wiring, which runs through the center of the rotator assembly. The two mounting plates for each configuration (parallel mount in Fig.~1(a) and perpendicular mount in Fig.~1(b)) were machined from cast epoxy MPC$4264$~\cite{MPP09}. An unassembled mPuck sample holder~\cite{CMR_mPuck} hosting the LCC$20$ package~\cite{SpectrumLCC20} with the quantum device under study, was integrated on top of these mounting plates.

The piezorotator, sample mount and sample holder are mechanically coupled to the mixing chamber of the dilution refrigerator, which has a nominal cooling power of $100~\mu$W at $100$~mK, via a $343$~mm long coldfinger with a diameter of $10$~mm, and a small `coldhead' as shown in Fig.~1. Both of these components were made from oxygen-free copper to maximize their thermal conductivity, and thereby dissipate the heat generated by the rotator mechanism as efficiently as possible. The coldfinger has two $3.5$~mm wide slots milled along its length to minimize eddy current heating. The entire rotation system is designed to be easily dismountable, unscrewing from the mixing chamber so that it can be replaced with a different experimental setup when necessary.

\noindent{\bf The Device Wiring:} Device measurements are typically conducted using very low excitation voltages ($\lesssim~100~\mu$V) or currents ($\lesssim~100$~nA) to minimize Joule heating. To avoid interference from the rotator control wires, which carry signals with higher frequencies and much larger amplitudes, the sample measurement wires follow a separate path down the dilution refrigerator to reach the mixing chamber. Fabric-woven wire loom containing twelve twisted pairs of $110~\mu$m diameter, $66~\Omega$/m (independent of temperature), $42$ gauge polyester insulated constantan wire~\cite{CMR_loom} runs from the room temperature Fischer connector down to the mixing chamber. This loom is heatsunk at the $4$~K plate, $1$~K pot plate and the mixing chamber. The remaining path from the mixing chamber to the device consists of a similar style of fabric-woven wire loom containing twelve twisted pairs of $90~\mu$m diameter, $2.3~\Omega$/m (at room temperature), $42$ gauge polyester insulated copper wire~\cite{CMR_loom}, which is soldered directly to the terminals of the mPuck sample holder. Photographs of the completed assembly with a device mounted in the rotator are presented in Fig.~2(a) and (c), which are top and side views, respectively.

\begin{figure}
\includegraphics[width=0.99\linewidth]{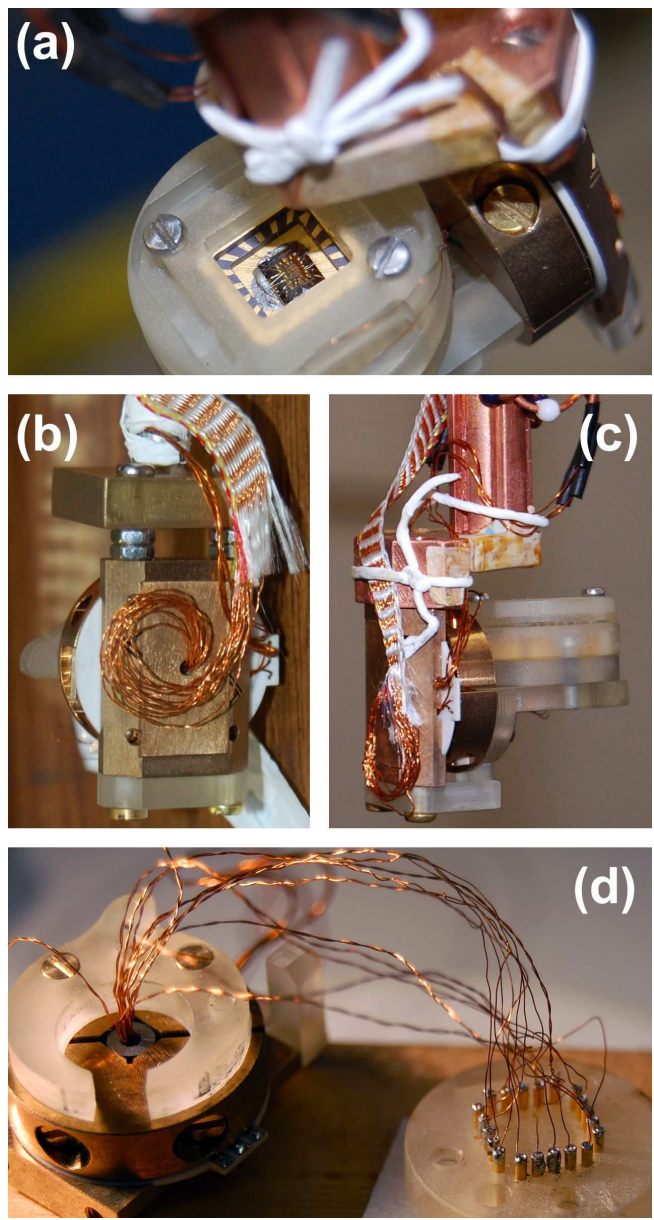}
\caption{Completed rotator assembly using the perpendicular mounting plate (i.e., the configuration in Fig.~1(b)). Top (a), back view showing the wiring loop (b) and side view (c) photographs are given. Part (d) shows the threading of sample loom through the center of the rotator after which it is soldered to the mPuck sample holder.}
\end{figure}

A major challenge in our design was how to achieve large rotation angles with twenty copper wires connected to the sample within the limited sample space. This issue was resolved by feeding the wires through a hole in the center axis of the piezo-rotator, as shown in Fig.~2(d) and soldering the ends into the pins of the mPuck sample holder base. To reduce tension in the wires upon rotation, a small spiral of free length is kept at the back of the rotator, as shown in Fig.~2(b). This design allows sufficient clearance for the sample loom to remain free to rotate with the sample, without coming into contact with the radiation shield and thermally shorting out the mixing chamber.

\noindent{\bf The Rotator Wiring:} Five wires run down to the rotator assembly, two of these supply the driving voltage to the piezo-element, the remaining three are used for the angular position sensor. The two drive lines require a very low resistance ($<~5~\Omega$) to ensure that the full drive voltage appears across the piezo, but at the same time, they need to have a low thermal conductivity to minimize heat leakage from the surroundings into the dilution refrigerator. The large driving signal for the piezo (up to $70$~V and up to $1$~kHz) means that these two wires need to be well shielded to prevent signal interference with the device and angle sensor lines. The angular sensor is a three-terminal resistance-based sensor with a total impedance of $20$~k$\Omega$. The series resistance of the wiring can be corrected for~\cite{SrinivasanICONN10} and so manganin wire can be used for these three sensor lines.

Three separate wiring segments were employed to meet these requirements. The first segment runs from a room-temperature Fischer connector down to the $4$~K plate. This segment consists of a home-made loom containing three insulated Cu wires (one twisted pair and one single wire) and four insulated manganin wires (two twisted pairs). The drive voltage passes through the Cu twisted pair, and the angle sensor uses three of the manganin wires. This loom passes down a  stainless-steel tube into the inner vacuum can (IVC), where it is heatsunk on the top-flange  and terminates at a single LEMO connector~\cite{LEMO}.

The second segment runs from the $4$~K plate to the mixing chamber, and consists of two parallel cryocables~\cite{Lakeshore07}, each containing four insulated cores made from superconducting NbTi wire with a braided, tinned copper outer shield and plastic cladding. One cryocable is used for the drive voltage signal and the other for the angle sensor to minimize cross-talk. The inner superconducting core minimizes electrical resistance, whilst maintaining low thermal conductivity. The plastic cladding is removed from both cryocables at the points at which it is clamped to the refrigerator, to ensure good thermalization and electrical grounding to the dilution unit.

The final segment runs from the mixing chamber to the rotator, where the two cryocables terminate at a LEMO connector and feed into five, oxygen-free copper coaxial cables~\cite{CoaxLtd} that run down the length of the coldfinger. The cores of these cables are soldered to the copper wires supplied with the rotator, for the final few centimeters into the rotator itself (see Fig.~2(c)). The shields of the coaxial cables are both grounded and thermalized via copper clamps at the top of the coldfinger.

\section{\label{sec:respow}ROTATOR POWER DISSIPATION DURING ACTUATION}

An important consideration for any low temperature {\it in situ} rotator is the amount of power dissipation during actuation. If this exceeds the available cooling power, low temperature operation cannot be maintained. Hence the first set of characterization measurements performed was that of power dissipation under continuous rotation at a variety of rotation speeds ranging from $0$ to $32~^{\circ}/$min, using the following procedure. For each set of rotation speeds, we record the mixing chamber temperature once it reaches equilibrium. This mixing chamber temperature is measured using a RuO$_{2}$ resistance thermometer mounted on the mixing chamber itself. The steady state mixing chamber temperature can be converted to an equivalent power dissipation through a separate calibration, performed with the rotator stationary. This calibration involves measuring the steady state mixing chamber temperature as a function of the power applied to a wire-wound heater on the mixing chamber. Using this calibration, we obtain the dissipation versus rotation speed data presented in Fig.~3.

It is important to note that the rotation speed is dependent upon both the amplitude and the frequency of the sawtooth driving voltage supplied to the piezo-element. For example, the same rotation rate can be achieved by using either less frequent but longer piezo strokes (i.e., higher amplitude, lower frequency driving voltage) or more frequent, but shorter piezo strokes, with the latter being the preferred option. Hence the data presented in Fig.~3 is sub-divided by the driving voltages applied for this test, with their corresponding frequencies displayed. Despite the different driving voltage/frequency combinations, the data in Fig.~3 fall onto a single curve, which indicates that the heat generated depends only upon the overall rotation speed. This strongly suggests that friction is the primary cause of heat dissipation in the rotator, which is expected given the nature of the inertial drive mechanism by which the rotator operates.

\begin{figure}
\includegraphics[width=0.99\linewidth]{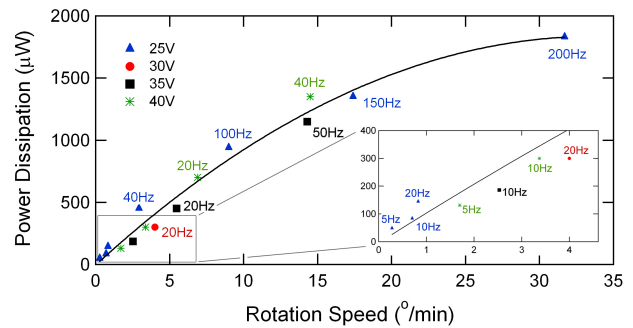}
\caption{Rotator power dissipation as a function of rotation speed, for a range of different amplitudes and frequencies of the drive voltage applied to the piezo-element. The insert on the right gives an expanded view of the lower frequency data in the curve.}
\end{figure}

The power dissipation of $<100~\mu$W for rotation speeds $<1^{\circ}$/min means that continuous rotation can be achieved without the mixing chamber exceeding $100$~mK. This performance is similar to the low temperature piezorotator system in Ref. \cite{OhmichiRSI01}, with an angle limited to $\pm10~^{\circ}$, as mentioned earlier. On average it takes around $20$~mins for a base temperature of $25$~mK to be recovered once rotation has stopped.

\section{\label{sec:angcal}ROTATION ACCURACY}

We now discuss the accuracy with which we can set and hold a given angle with our low-temperature rotator. As mentioned earlier, angular read-out is achieved using a three-terminal resistance-based sensor with a total impedance of approximately $20$~k$\Omega$. The sensor is essentially a rheostat, consisting of a film with fixed resistivity mounted on the rotator base plate, and a third terminal mounted on the rotator that slides along the film as the angle is changed. With a voltage $V_{0}$ across the film, the voltage $V$ at the third terminal is used to obtain the angle, varying linearly between $0$~V at $0^{\circ}$ and $V_{0}$ at $306^{\circ}$ with a small dead region where no angle can be read. A small correction for the series resistance of the wiring to the sensor needs to made, which is detailed elsewhere~\cite{SrinivasanICONN10}.

\begin{figure}
\includegraphics[width=0.99\linewidth]{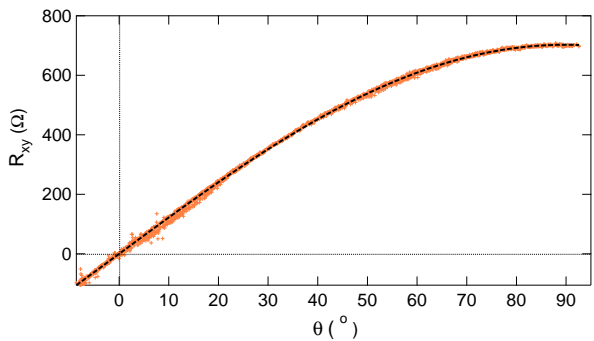}
\caption{The measured Hall resistance $R_{xy}$ of the 2DEG in a GaAs HEMT versus the angle $\theta$ measured by the resistive angle sensor. The dashed black line is a sinusoidal fit to the data. The quality of the fit demonstrates the accuracy of the resistive angle sensor used to measure the rotator position.}
\end{figure}

One concern was the readout accuracy at low temperatures, so a series of independent measurements were taken to verify the angle readout of the rotator. An AlGaAs/GaAs High Electron Mobility Transistor (HEMT) containing a two-dimensional electron gas (2DEG) \cite{wafer}, was used to conduct Hall effect measurements as a function of angle whilst being rotated continuously at $4~^{\circ}$/min through $90^{\circ}$ between the parallel and perpendicular device orientations, relative to a fixed magnetic field of $B = 0.15$~T. From the measured Hall data, the angles could be independently calculated and compared against the controller readout angle values in order to generate a calibration curve.

In Fig.~4 we show the measured Hall resistance $R_{xy}$ versus the angle $\theta$ measured by the rotator sensor, corrected for series resistance and offset such that $0~^{\circ}$ coincides with $R_{xy} = 0~\Omega$. The data (orange) was obtained by using a lock-in amplifier to continuously monitor the Hall voltage $V_{xy}$ with a constant excitation current of $I = 10$~nA at $19$~Hz flowing through the 2DEG, during a single $90~^{\circ}$ rotation at a rate of $31.7~^{\circ}/$min. The measured $R_{xy}$ in Fig.~4 should exhibit a sinusoidal dependence on $\theta$ if the angle is being read correctly over the entire range. Such a sinusoidal fit to the data is presented as the dashed black line in Fig.~4, with the only free parameter being the peak Hall resistance $R_{max} = 702~\Omega$.

Another consideration is the accuracy with which a fixed angle can
be held over an extended length of time. This can be achieved in two
possible ways; open-loop mode in which the rotator relies solely on
friction to remain stationary, and closed-loop mode where the
feedback loop in the controller continually monitors the angle and
actively corrects for any deviations from the desired angle by
actuating the rotator.

The open-loop configuration is the most desirable of the two as it
produces the least power dissipation and thus the lowest base
temperature of $24$~mK, however the angular stability over time
required some investigation. We initially performed a `set and
settle' test at zero magnetic field, which involved moving the
rotator to the target angle, holding it there briefly in closed-loop
mode before switching to open-loop mode and monitoring the angle for
three minutes. The measurement was repeated three times and in each
case there was an initial drift in angle, always in the same
direction with a magnitude of $\sim1^{\circ}$. After this initial
settling of the rotator, the angle remains stable to within $\sim
\pm 0.015^{\circ}$ over extended periods of at least $20$~min.
Similar results are obtained when performed at other angles and the
reproducible nature of this settling offset means that it can be
corrected for, if desired.

We also examined the stability of the rotator as a function of applied magnetic field. In the first series of measurements the 2DEG sample was used, with the following measurement protocol: The desired angle was set in closed loop mode, then the controller was switched to open-loop mode, and the magnetic field was swept from $B = 0$~T up to $2$~T and back down to $0$~T at a rate of $0.1$~T/min over a total time of $40$~min. This duration is much greater than the $\sim1$~min settling time, hence any change in angle over this period is independent of time drift. This test was repeated for different angles and it was found that there was a slight linear change in angle with magnetic field that was dependent upon the target angle. The maximum change was just under $\sim0.05~^{\circ}$/T from the initial settling position at $180.6^{\circ}$. In order to reduce this change of angle with magnetic field, a second set of measurements were performed with a 2D hole sample held within a LCC20 leadless chip carrier sourced from a different supplier. For these measurements, the magnetic field was swept
from -10T to +10T and back at a rate of $0.2$T/min. The rotator angle readout showed a change of $\sim0.02~^{\circ}$/T between 0T and $\pm10$T which is much less than the previous measurements given the wider field range. One possibility for this improved stability may be that the LCC20 chip carrier holding the first sample, may have more magnetic components (such as a Nickel flashing) which causes a torque on the rotator in an applied field.

\begin{figure}
\includegraphics[width=0.99\linewidth]{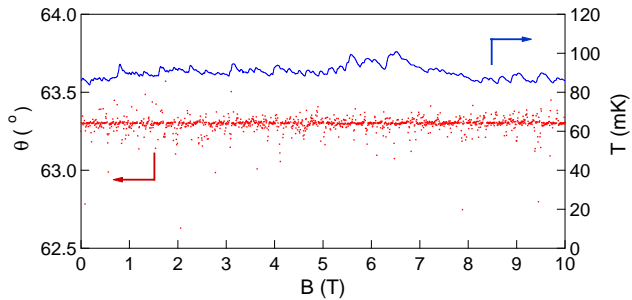}
\caption{A plot of rotator position (red data/left axis) measured with the angle sensor as a function of applied magnetic field $B$ at fixed angle in closed-loop mode. The corresponding mixing chamber temperature $T$ (blue data/right axis) is also presented. The magnetic field is swept from $0$ to $10$~T at a rate of $0.4$~T/min such that the complete trace takes $25$~mins.}
\end{figure}

To ensure that there is no change in angle with magnetic field, even if the sample or package has magnetic elements, closed-loop mode operation can be used to keep the angle fixed at a given target. To demonstrate this, the measurements were repeated with the 2DEG sample in closed-loop mode
between $0$~T and $10$~T, shown in Fig.~5. With active feedback enabled, the angle can be held constant to within $\pm0.03^{\circ}$ over the full field range. Although the trade-off is an increase in base temperature, the
mixing chamber temperature remains below $100$~mK, averaging around
$90$~mK.

\section{\label{sec:QHEdata}DEVICE MEASUREMENT}

Finally, we present quantum Hall effect measurements made with the rotator in closed-loop mode to demonstrate that conditions suitable for observing quantum transport effects such as spin-splitting can be achieved, despite the additional heat-load generated by active feedback control of the rotator. Figure~6 is a graph of $R_{xx}$ versus perpendicular magnetic field $B_{\perp}$ for angles of $\theta = 22.0^{\circ}$, $14.5^{\circ}$ and $7.2^{\circ}$. For each trace the total field $B$, was swept from $0$~T to $8$~T. These measurements were performed with the rotator in the configuration depicted in Fig.~2, with $\theta = 90^{\circ}$ corresponding to the 2DEG plane perpendicular to the magnetic field and $\theta = 0^{\circ}$ corresponding to an in-plane field. The longitudinal resistance $R_{xx}$ was measured using a lock-in amplifier with a constant excitation current of $10$~nA at a frequency of $19$~Hz.

\begin{figure}
\includegraphics[width=0.99\linewidth]{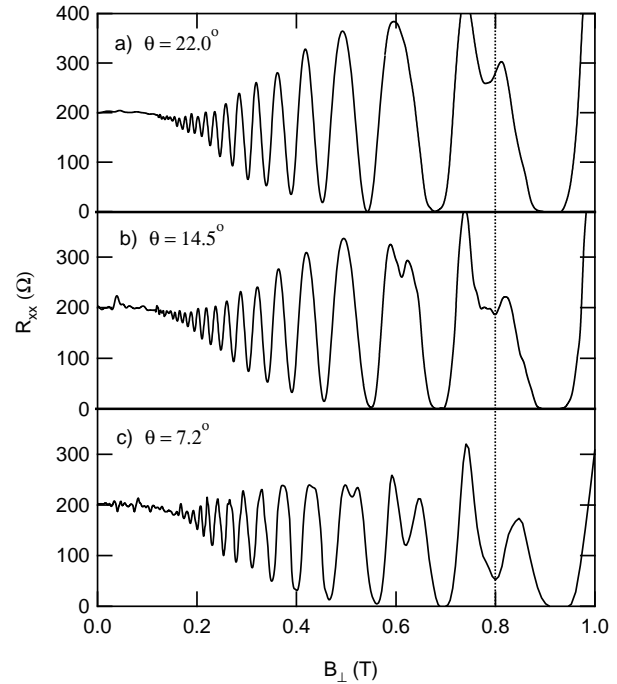}
\caption{The measured longitudinal resistance $R_{xx}$ from the 2DEG in a GaAs HEMT as a function of perpendicular magnetic field component $B_{\perp} =$ B cos $\theta$ for three different tilt angles (a) $\theta = 22.0^{\circ}$, (b) $\theta = 14.5^{\circ}$ and (c) $\theta = 7.2^{\circ}$. The angle $\theta$ is the angle between the applied magnetic field $B$ and the plane of the 2DEG. Although the $B_{\perp}$ range is common in all three panels, it is achieved with different ranges of B. The dashed vertical line where $B_{\perp} = 0.8$~T corresponds to B $= 2.14$, $3.20$ and $6.38$~T respectively for parts (a) to (c). Measurements were made with the rotator in closed-loop mode with a mixing chamber temperature of $T \sim 90$~mK.}
\end{figure}

Each trace exhibits Shubnikov - de Haas (SdH) oscillations, arising from Landau quantization, which are periodic in $1/B_{\perp}$. The key variations between each of the three traces is most apparent in the maxima centered around $B_{\perp} = 0.53$, $0.62$ and $0.8$~T. In the top trace, where $\theta = 22.0~^{\circ}$, the total field is relatively small, hence Zeeman spin-splitting of the Landau levels is only just observable in the SdH maxima at $B_{\perp} = 0.8$~T. At smaller angles, closer to the in-plane field orientation, a larger total field is required to obtain the same SdH maxima and minima points. Hence the Zeeman splitting for a given perpendicular field component increases as $\theta$ decreases. This can be seen in the bottom trace, which is fully spin resolved and exhibits two distinct peaks, in comparison to the top trace which has a larger $\theta$. Thus the rotator allows us to observe spin-splitting in the SdH maxima at lower $B_{\perp}$ values.

This observation provides the final proof-of-concept that the rotator can be used to study quantum effects which require {\it in situ} tilting with respect to a magnetic field.

\section{\label{sec:conc}CONCLUSION}

In conclusion we have reported the design and development of a  piezoelectric sample rotation system, and its integration into an Oxford Instruments Kelvinox $100$ dilution refrigerator, for orientation-dependence studies of quantum transport in semiconductor nanodevices at millikelvin temperatures in magnetic fields up to $10$~T. Our apparatus allows for continuous {\it in situ} rotation of a device through $>100^{\circ}$ in two possible configurations. The first enables rotation of the field within the plane of the device and the second allows the field to be rotated from in-plane to perpendicular to the device plane. The rotation system has a power dissipation less than $2$~mW for high rotation speeds $\sim30~^{\circ}$/min, but we can achieve a power dissipation less than $100~\mu$W if we rotate slowly $< 1~^{\circ}/$min. In closed-loop mode, the rotator can maintain an angle to within $\pm0.03^{\circ}$ over a magnetic field range of $0 < B < 10$~T with a mixing chamber temperature of $\sim90$~mK. A lower temperature can be achieved with active feedback turned off and the post rotation settling corrected for.

\begin{acknowledgments}
The authors wish to acknowledge J. Cochrane for technical support and L.H. Ho, D.A. Ritchie, P. Atkinson and M. Pepper for providing the GaAs HEMT sample used for Figs.~4, 5 and 6. We would also like to thank Markus Janotta at Attocube Systems AG and Thomas Ihn of ETH Zurich for helpful advice. This work was supported by Australian Research Council (ARC) Grants No. DP$0772946$ and DP$0986730$. APM and ARH acknowledge ARC Future and Professorial Fellowships, respectively.
\end{acknowledgments}


\begin{thebibliography}:

\bibitem{WolfSci01} S.A. Wolf, D.D. Awschalom, R.A. Burhman, J.M. Daughton, S. von Moln\'{a}r, M.L. Roukes, A.Y. Chtchelkanova, and D.M.S.A. Treger, Science {\bf 294}, 1488 (2001).

\bibitem{AwschalomPhys09} D.D. Awschalom and N. Samarth, Physics {\bf 2}, 50 (2009).

\bibitem{WinklerBook03} R. Winkler, {\it Spin-orbit coupling effects in two-dimensional electron and hole systems}, (Springer Tracts in Modern Physics, Vol. 191, Springer, Berlin, 2003).

\bibitem{DanneauPRL06} R. Danneau, O. Klochan, W.R. Clarke, L.H. Ho, A.P. Micolich, M.Y. Simmons, A.R. Hamilton, M. Pepper D.A. Ritchie, and U. Z\"{u}licke, Phys. Rev. Lett. {\bf 97}, 026403 (2006).

\bibitem{KlochanNJP09} O. Klochan, A.P. Micolich, L.H. Ho, A.R. Hamilton, K. Muraki and Y. Hirayama, New. J. Phys. {\bf 11}, 043018 (2009).

\bibitem{ChenNJP10} J.C.H. Chen, O. Klochan, A.P. Micolich, A.R. Hamilton, T.P. Martin, L.H. Ho, U. Z\"{u}licke, D. Reuter and A.D. Wieck, New. J. Phys. {\bf 12}, 033043 (2010).

\bibitem{BoebingerPRL90} G.S. Boebinger, G. Montambaux, M.L. Kaplan, R.C. Haddon, S.V. Chichester and L.Y. Chiang, Phys. Rev. Lett. {\bf 64}, 591 (1990).

\bibitem{OsadaPRL91} T. Osada, A. Kawasumi, S. Kagoshima, N. Miura and G. Saito, Phys. Rev. Lett. {\bf 66}, 1525 (1991).

\bibitem{StorrPRB91} K. Storr, L. Balicas, J.S. Brooks, D. Graf and G.C. Papavassiliou, Phys. Rev. B {\bf 64}, 045107 (2001).

\bibitem{RaffyPRL91} H. Raffy, S. Labdi, O. Laborde and P. Monceau, Phys. Rev. Lett. {\bf 66}, 2515 (1991).

\bibitem{HoferPRB00} J. Hofer, T. Schneider, J.M. Singer, M. Willemin, H. Keller, T. Sasagawa, K. Kishio, K. Conder and J. Karpinski, Phys. Rev. B {\bf 62}, 631 (2000).

\bibitem{MaoPRL00} Z.Q. Mao, Y. Maeno, S. Nishizaki, T. Akima and T. Ishiguro, Phys. Rev. Lett. {\bf 84}, 991 (2000).

\bibitem{EisensteinPRB90} J.P. Eisenstein, H.L. St\"{o}rmer, L.N. Pfeiffer and K.W. West, Phys. Rev. B {\bf 41}, 7910 (1990).

\bibitem{SchmellerPRL95} A. Schmeller, J.P. Eisenstein, L.N. Pfeiffer and K.W. West, Phys. Rev. Lett. {\bf 75}, 4290 (1995).

\bibitem{MurakiPRB99} K. Muraki and Y. Hirayama, Phys. Rev. B {\bf 59}, R2502 (1999).

\bibitem{KumadaPRL02} N. Kumada, D. Terasawa, Y. Shimoda, H. Azuhata, A. Sawada, Z.F. Ezawa, K. Muraki, T. Saku and Y. Hirayama, Phys. Rev. Lett. {\bf 89}, 116802 (2002).

\bibitem{MeierJMMM00} G. Meier, D. Grundler, K.-B. Broocks, Ch. Heyn and D. Heitmann, J. Magn. Magn. Mater. {\bf 210}, 138 (2000).

\bibitem{AbePRL07} N. Abe, K. Taniguchi, S.Ohtani, T. Takenobu, Y. Iwasa and T. Arima, Phys. Rev. Lett. {\bf 99}, 227206 (2007).

\bibitem{MansonCM08} J.L. Manson, M.M. Conner, J.A. Schleuter, A.C. McConnell, H.I. Southerland, I. Malfant, T. Lancaster, S.J. Blundell, M.L. Brooks, F.L. Pratt, J. Singleton, R.D. McDonald, C. Lee and M.-H. Whangbo, Chem. Mater. {\bf 20}, 7408 (2008).

\bibitem{BhattacharyaRSI98} A. Bhattacharya, M.T. Tuominen and A.M. Goldman, Rev. Sci. Inst. {\bf 69}, 3563 (1998).

\bibitem{PalmRSI99} E.C. Palm and T.P. Murphy, Rev. Sci. Inst. {\bf 70}, 237 (1999).

\bibitem{deHaasRSI96} E. de Haas, W. Barsingerhorn and J.F. van der Veen, Rev. Sci. Inst. {\bf 67}, 1930 (1996).

\bibitem{OhmichiRSI01} E. Ohmichi, S. Nagai, Y. Maeno, T. Ishiguro, H. Mizuno and T. Nagamura, Rev. Sci. Inst. {\bf 72}, 1914 (2001).

\bibitem{wafer} These 2DEG devices were fabricated from the A$2899$ wafer grown by Cambridge.

\bibitem{SpectrumLCC20} The LCC20 packages were ordered from Spectrum Semiconductor Materials, Inc. The part number is LCC02034 and the manufacturer's drawing number is IRK20F1-5856B.

\bibitem{AttoRot09} Attocube Systems AG, Rotator Type: ANRv$51$/LT/UHV/RES.

\bibitem{MPP09} MPC $4264$ amber is an unfilled, anhydride cured, cast epoxy resin made by Maryland Plastic Products, http://www.marylandplasticprdts.com/castepoxy.html

\bibitem{CMR_mPuck} Two separate pieces were ordered from Cambridge Magnetic Refrigeration: CMR mP-body $12$CU-LF$24$, Part No. $02-26-029$ without the copper loom and CMR mP-clamp, Part No. $02-26-027$ and assembled to form the whole sample holder.

\bibitem{CMR_loom} Ordered from the CMRdirect 2007 Catalogue; the Constantan loom model number is: $02-32-002$(CMR/CWL-$12$CO) and the Copper loom model number is: $02-32-003$ (CMR/CWL-$12$Cu).

\bibitem{SrinivasanICONN10} A. Srinivasan, L.A. Yeoh, T.P. Martin, O. Klochan, A.P. Micolich and A.R. Hamilton, Proceedings of the 2010 International Conference on Nanoscience and Nanotechnology ($ICONN-2010$), IEEE Conf. Proc. (In Press).

\bibitem{LEMO} Non-metallic LEMO connector part numbers used are: EGG.1B.308.ZLL and FGG.1B.308.ZLA.

\bibitem{Lakeshore07} Lake Shore Cryotronics, Inc., Cryogenic Cable Type: CryoCable$^{TM}$ Type: CYRC (2007).

\bibitem{CoaxLtd} The oxygen-free copper coaxial cables were ordered from Coax Co., Ltd.; Part number SC-$219/50$ with an outer diameter of $2.19$~mm and a central conductor of diameter $0.510$~mm.

%
%













\end{thebibliography}
\end{document}